\documentclass[pdflatex,sn-mathphys-num]{sn-jnl}
\usepackage{graphicx}%
\usepackage{multirow}%
\usepackage{amsmath,amssymb,amsfonts}%
\usepackage{amsthm}%
\usepackage{mathrsfs}%
\usepackage[title]{appendix}%
\usepackage{xcolor}%
\usepackage{textcomp}%
\usepackage{manyfoot}%
\usepackage{booktabs}%
\usepackage{algorithm}%
\usepackage{algorithmicx}%
\usepackage{algpseudocode}%
\usepackage{listings}%
\usepackage{svg}
\usepackage{caption}
\usepackage{fullpage}
\usepackage{pgffor}
\usepackage{rotating}
\usepackage{hyperref}
\usepackage[normalem]{ulem} 
\theoremstyle{thmstyleone}%

\theoremstyle{thmstyletwo}%
\theoremstyle{thmstylethree}%

\usepackage{float}
\floatstyle{plain}
\newfloat{extfigure}{htb}{lop}
\floatname{extfigure}{\textbf{Extended Data Fig.}}
\usepackage{caption}
\begin{document}

\title[As Cities Grow, They Spread Out Rather Than Rise]{As Cities Grow, They Spread Out Rather Than Rise}

\author[1,2]{\fnm{Peiran} \sur{Zhang}}
\author[3]{\fnm{Fabiano L.} \sur{Ribeiro}}
\author*[1,4]{\fnm{Liang} \sur{Gao}}%
\author[3]{\fnm{Hudson Vinícios Tavares} \sur{Mineiro}}
\author[1,4]{\fnm{Bin} \sur{Jia}}
\author*[1,4]{\fnm{Ziyou} \sur{Gao}}%
\author[5]{\fnm{Michael \sur{Batty}}}
\author*[2,6]{\fnm{Diego} \sur{Rybski}} %

\affil[1]{\orgdiv{School of Systems Science}, \orgname{Beijing Jiaotong University}, \orgaddress{\city{Beijing}, \postcode{100044}, \country{China}}}

\affil[2]{\orgdiv{Research Area Spatial Information and Modelling}, \orgname{Leibniz Institute of Ecological Urban and Regional Development (IOER)}, \orgaddress{\street{Weberplatz 1}, \city{Dresden}, \postcode{01217}, \country{Germany}}}

\affil[3]{\orgdiv{Department of Physics (DFI)}, \orgname{Federal University of Lavras (UFLA)}, \orgaddress{\city{Lavras}, \postcode{37200-900}, \state{MG}, \country{Brazil}}}

\affil[4]{\orgdiv{Key Laboratory of Transport Industry of Big Data Application Technologies for Comprehensive Transport}, \orgname{Beijing Jiaotong University}, \orgaddress{\city{Beijing}, \postcode{100044}, \country{China}}}

\affil[5]{\orgdiv{Centre for Advanced Spatial Analysis}, \orgname{University College London}, \orgaddress{\street{149 Tottenham Court Road}, \city{London}, \postcode{W1T 5AU}, \country{United Kingdom}}}

\affil[6]{\orgname{Complexity Science Hub (CSH)}, \orgaddress{\street{Metternichgasse 8}, \city{Vienna}, \postcode{1030}, \state{Vienna}, \country{Austria}}}


\abstract{
Cities and settlements are a global phenomenon, and their continued expansion is fundamentally transforming patterns of resource demand.
This transformation is reflected in growing pressures on land, material use, and infrastructure.
Thus, understanding how cities grow in space is essential for planning future urban development and determining all its consequences.
Here, we characterize four geometrically related dimensions of urban growth: two horizontal dimensions describing the urban area, one vertical dimension represented by average building height, which together with the urban area defines building volume, and the population dimension.
Analyzing a global dataset of growing cities, for the period 1975-2025 we identify three main patterns.
First, urban growth is fundamentally anisotropic -- as cities gain population, they expand much more rapidly in the horizontal than in the vertical direction.
Second, average building height decreases in the vast majority of cities as they grow, indicating that vertical development is not only slower than horizontal expansion but on average, can act in an effectively negative manner by giving greater weight to the horizontal dimensions.
Third, despite their diverse growth trajectories, cities converge toward characteristic population densities, both with respect to urban area and building volume.
We further explore alternative scenarios for future urban growth. These projections show that different assumptions about horizontal and vertical scaling can lead to substantially different demands for urban land and building volume.
These findings reveal constraints on the long-term evolution of urban form and provide a quantitative assessment for understanding the spatial dynamics and development of cities. 
}

\maketitle

\section{Introduction}\label{sec1}
Cities either expand outward or upward~\cite{batty2018inventing,mahtta2019building}. 
Horizontal expansion has accompanied urban development since the earliest settlements~\cite{longley1991size,ortman2015settlement}. By contrast, vertical development emerged as a complementary mode of urban growth only after the invention of steel-frame construction and elevators, which made increasingly tall buildings both technically feasible and functionally practical~\cite{SullivanL1896,BuitenhuisP1957,ahlfeldt2022economics}. Today, this shift from purely horizontal growth toward a combination of horizontal and vertical expansion is evident in cities of many sizes worldwide~\cite{frolking2024global}. Although not universal, vertical growth has become a defining feature of the modern
cityscape and is generally associated with more compact and concentrated urban forms~\cite{li2026global}.
Understanding how cities balance horizontal and vertical expansion is important because these distinct growth modes shape fundamental urban characteristics, including land consumption \cite{JehlingKBR2025}, infrastructure demand~\cite{prieto2025urban}, energy use~\cite{zhou2022satellite,batik2026best}, 
and the intensity of socio-economic interactions \cite{Schlapfer2014,Ghafoor2019,Alencar2025}. In addition, the interplay between outward and upward growth influences urban density and, consequently, the mobility, efficiency, and productivity of cities. Despite its importance, the mechanisms that govern the relationship between population growth, building area,  urban form, and building height are not yet sufficiently understood, particularly on a global scale.

The population represents a natural reference for understanding urban evolution. As the most widely used indicator of city size, it offers an intuitive basis for comparing cities and helps reduce differences that arise simply from variations in absolute scale. Following this logic, previous studies have proposed urban sprawl indices, which relate area growth to population growth~\cite[e.g.]{fulton2001sprawls,BurkeM2002}, while more recent work has extended this approach to three-dimensional urban expansion by incorporating building volume~\cite{hou2026disparities}. 
However, these measures are typically calculated using only an early and a later state of urban development, yielding a static indicator rather than a generalized long-term relationship between population growth and the evolving three-dimensional form of cities.

The paper in hand complements publications in the urban scaling context in two respects.
First, in addition to relating area and population, it includes volume.
Second, instead of an cross-sectional analysis (transversally) the analysis is performed across time (longitudinally).
Accordingly, we can actually speak about \emph{growth} when relating area, volume, and population.
That is, three spatial dimensions and the population dimension, together characterizing the four dimensions of urban growth.

Urban scaling theory describes urban indicators as power-law functions of population size ($P$)~\cite{bettencourt2007growth,lobo2020settlement,ribeiro2023mathematical} mostly in a cross-sectional manner, i.e., comparing a set of cities at the same time.
For an urban quantity ($Y$), the relationship is expressed as $Y \sim P^\beta$, where $\beta$ is the scaling exponent. Values of $\beta$ are commonly interpreted as superlinear ($\beta > 1$) for socioeconomic outputs, linear ($\beta \approx 1$) for basic individual needs, and sublinear ($\beta < 1$) for infrastructure-related quantities~\cite{bettencourt2013origins}. 
The urban area is a well-established example of sublinear scaling, implying that area increases less than proportionally with population.
Consequently, larger cities tend to be denser on average, a pattern closely related to urban allometry~\cite{nordbeck1971urban,ribeiro2025buildings}.
In contrast, much less is known about the scaling behavior of cities' vertical form. 
Less than a handful of studies have examined how building height scales with population size. For example, \citet{schlapfer2015skyline} adapted the urban scaling framework to investigate the static cross-sectional relationship between average building height and population across 12 North American cities. More recently, additional studies have explored building-height scaling in specific countries and regions~\cite[e.g.]{Molinero2021}, but evidence remains fragmented and geographically limited. Overall, while the scaling of horizontal urban expansion has been extensively documented, the scaling properties of vertical development remain poorly understood, particularly from a three-dimensional perspective that jointly considers urban area, building height, and population size.

Urban growth is inherently dynamic, whereas cross-sectional scaling captures statistical regularities across cities at a given point in time rather than the developmental trajectories of individual cities~\cite{bettencourt2020interpretation,ribeiro2020relation,marquis2026meaning}. Consequently, cross-sectional scaling exponents cannot fully reveal how a particular city evolves over time. Despite the extensive literature on cross-sectional scaling, longitudinal scaling has received considerably less attention and has been investigated primarily through case studies of a limited number of cities~\cite[e.g.]{depersin2018global,keuschnigg2019scaling,gravier2024typology}.
Existing evidence suggests that temporal scaling relationships can differ substantially from their cross-sectional counterparts. For example, some cities, such as Shanghai (China), exhibit superlinear scaling between urban area and population over time, in contrast to the sublinear relationship commonly observed across cities~\cite{xu2020scaling}. This discrepancy suggests that the greater density of larger cities observed in cross-sectional analyses does not necessarily imply that individual cities become denser as they grow. A longitudinal scaling analysis is therefore needed to clarify how horizontal and vertical urban expansion co-evolve with population growth, and to determine whether the scaling patterns observed across growing cities are consistent with the developmental trajectories of individual cities.

\section{Results}
In the following, we investigate the scaling relationships among population ($P$, in persons), residential building area ($A$, in m$^2$), and residential building volume ($V$, in m$^3$) for 589 cities worldwide. As described in the Materials and Methods Sec.~\ref{sec:mat-methods}, our analysis is restricted to the area and volume of residential buildings, as these primarily reflect the spatial demand associated with accommodating the population. Furthermore, because our focus is on growing cities, we include only those that exhibited continuous population growth over the period 1975–2025.

\begin{figure}[H]
\centering
\includegraphics[width=0.75\linewidth]{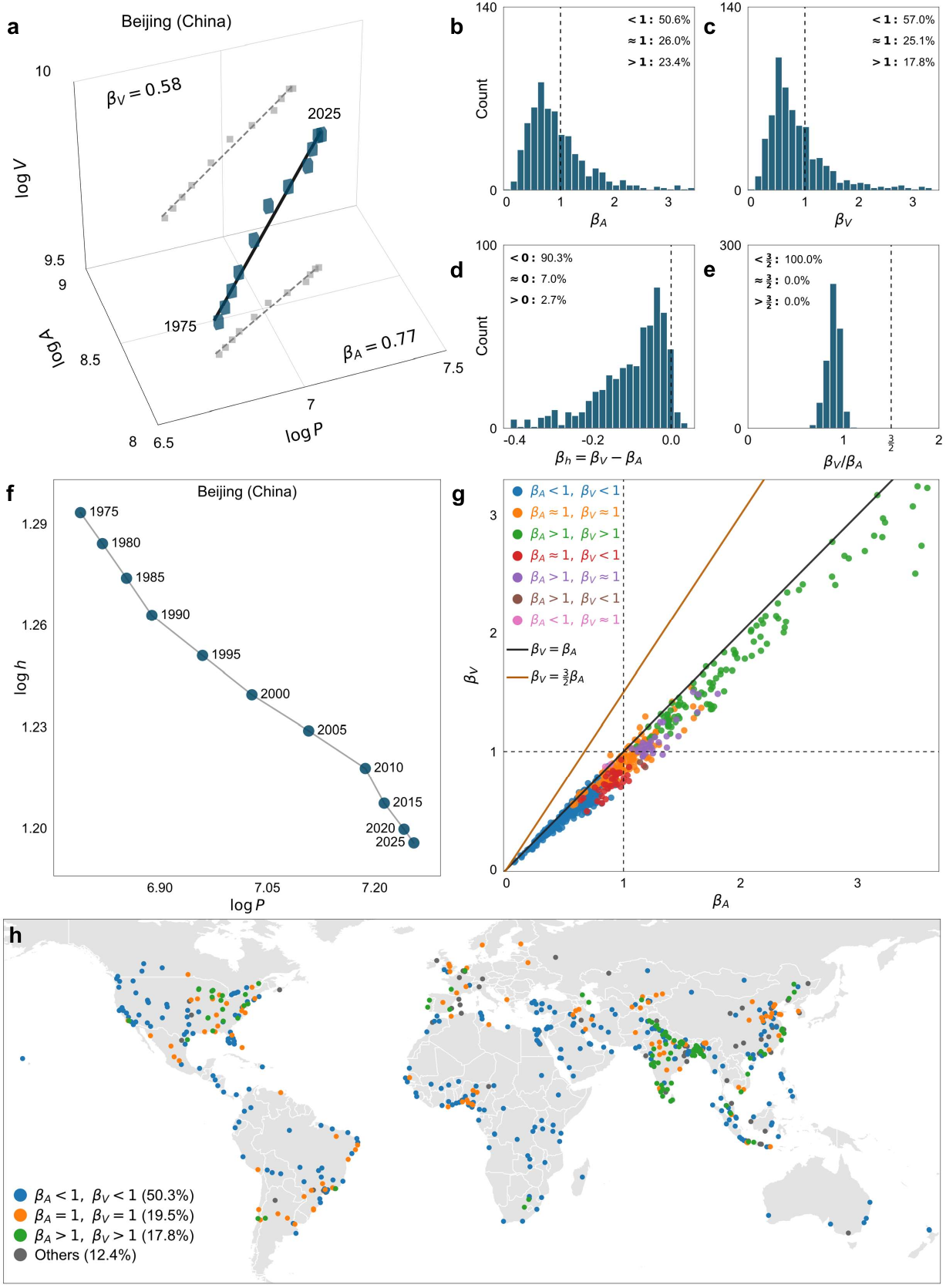}
\caption{\textbf{Scaling of population, area, and volume of growing global cities.}
(a) For the example city Beijing (China), the three quantities population ($P$), residential building area ($A$), and residential building volume ($V$) are located in 3D-scatter plot together with the regression obtained from PCA (all axes in log-scale).
Each dot represents a year (from 1975 to 2025; 5\,y steps).
From the 2D projections (gray), we obtain the individual scaling relations between pairs of these quantities, in particular $\beta_A$ and $\beta_V$, Eqs.~(\ref{Eq_AP}) and~(\ref{Eq_VP}), respectively.
See \href{https://zhangpprr.github.io/Urban-temproal-scaling-of-3D-forms/}{weblink} for an interactive visualization of all the cities considered.
(b,c,d) Histograms of the three scaling exponents $\beta_A$, $\beta_V$, $\beta_h$, where the last is obtained from Eq.~(\ref{Eq_betah}).
The fact that $\beta_h$ is typically smaller than 0 means that the average building height typically decreases with growing population.
(e) Histogram of $\beta_V/\beta_A$.
The vertical dashed line is located at 3/2 and indicates isotropic scaling.
(f) For the example city, the average building height is plotted vs.\ its population (in log-scale) between 1975 (left) and 2025 (right).
As expected from (d), $\beta_h<0$ for most cities, and also in this case, the average building height is decreasing.
(g) Plot of $\beta_V$ vs.\ $\beta_A$ for all analyzed cities.
Each dot represents a city and is colored according to the respective category defined by the different scaling behavior (exponents of area and volume being smaller or larger than 1 (at 99\,\% confidence), or equal, indicated by the dashed horizontal and vertical lines).
The uncertainty is obtained from 5,000 bootstrap realizations. 
The black diagonal line represents the case $\beta_V=\beta_A$, and the brownish line is given by $\beta_V=\frac{3}{2}\beta_A$ (isotropic scaling). 
(h) Geographical position of all analyzed cities with colors according to the categories used in panel (g). 
The percentages in parentheses indicate the proportion of cities belonging to each specific category.
}
\label{fig:VAP}
\end{figure}

Figure~\ref{fig:VAP}a illustrates an example of scaling relationships for Beijing (China). 
The trajectory captures the city's evolution from low (1975) to high (2025) values of population, area, and volume.
In logarithmic space, this growth follows an approximately linear trend, allowing us to fit a three-dimensional regression using Principal Component Analysis (PCA).
The projections onto the coordinate planes correspond to the scaling relations commonly studied in urban scaling, namely $A\sim P^{\beta_A}$ and $V\sim P^{\beta_V}$, see Eqs.~(\ref{Eq_AP}) and~(\ref{Eq_VP}). The former is known as the fundamental allometry \cite{nordbeck1971urban,batty2011defining,burger2022global,ribeiro2023mathematical}. For Beijing, both scaling exponents are clearly smaller than one, indicating that urban area and building volume increase more slowly than population.

Extracting the scaling exponents for all considered cities, we obtain the distributions shown in Fig.~\ref{fig:VAP}b and~c. For the area scaling exponent, we find that, at the 99\,\% confidence level, 50.6\,\% of cities have $\beta_A<1$, 26.0\,\% are statistically indistinguishable from $\beta_A=1$, and 23.4\,\% have $\beta_A>1$.
These results are qualitatively consistent with previous cross-sectional studies \cite[e.g.,][]{burger2022global}.
The distribution of the volume scaling exponent exhibits a similar pattern (Fig.~\ref{fig:VAP}c).

Since building volume is given by the product of urban area and average building height, a third scaling relation follows, $h \sim P^{\beta_h}$, Eq.~(\ref{Eq_hP}), %
and the corresponding scaling exponent satisfies $\beta_h=\beta_V-\beta_A$, Eq.~(\ref{Eq_betah}).
The distribution of $\beta_h$ is shown in Fig.~\ref{fig:VAP}d. In contrast to the area and volume exponents, $\beta_h$ is not only smaller than one but is negative for more than 90\,\% of the cities ($\beta_h<0$). This indicates that in most cities, the average building height decreases as the population increases.
This, of course, does not imply that individual buildings become shorter over time. Rather, urban growth is predominantly accommodated by the construction of new buildings -- typically located at the urban fringe -- that are, on average, shorter than the existing building stock. As a result, the mean building height decreases despite continued urban development.
This behavior is illustrated in Fig.~\ref{fig:VAP}f for the case of Beijing, where the average building height is plotted against population. As the city grows over time, the average building height exhibits a steady decline.

Figure~\ref{fig:VAP}g summarizes these findings by plotting $\beta_V$ against $\beta_A$ for all cities in our sample. The two exponents are strongly correlated, forming an elongated point cloud located just below the diagonal ($\beta_V=\beta_A$). This location implies $\beta_h<0$ for the majority of cities, consistent with the distribution shown in Fig.~\ref{fig:VAP}d, and therefore indicates a decreasing average building height.
For comparison, we also include the reference line $\beta_V=\frac{3}{2}\beta_A$, which corresponds to isotropic growth. Under this scenario, cities would expand equally along the two horizontal dimensions and the vertical dimension. The empirical exponents, however, lie well below this line, indicating that urban expansion is far from isotropic.
This conclusion is further supported by Fig.~\ref{fig:VAP}e, which shows the distribution of the exponent ratio $\beta_V/\beta_A$. Under isotropic growth, this ratio would be equal to $3/2$, whereas the observed values are predominantly smaller than one. Interestingly, the distribution of $\beta_V/\beta_A$ is approximately Gaussian. Together, these results indicate that cities expand primarily horizontally and, in relative terms, even contract vertically.

It is important to place our classification within the context of previous studies that have also sought to characterize cities according to their growth patterns. For instance,~\citeauthor{mahtta2019building} classified cities according to the relative contributions of upward and outward urban growth. Similarly,~\citeauthor{zhou2022satellite} developed a classification based on population density and building height. These approaches complement our scaling-based classification and illustrate the diversity of ways in which the horizontal and vertical dimensions of urban growth can be characterized. For a recent review of the spatial growth of cities and the different approaches used to describe urban expansion, see \cite{Marquis2025}.

\begin{figure}[H]
\centering
\includegraphics[width=1.0\linewidth]{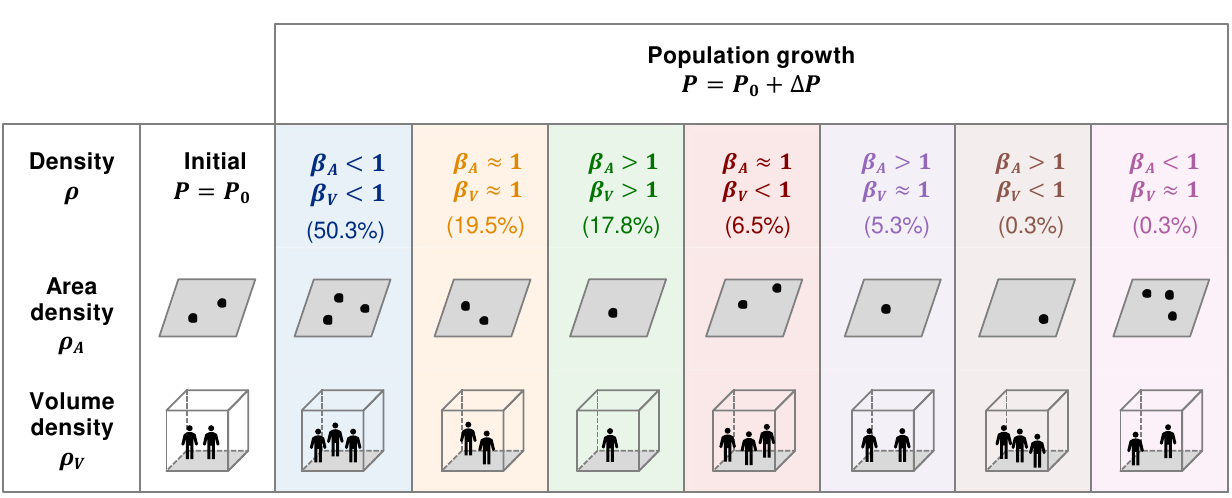}
\caption{
Illustration of how the area and volume exponents affect the surface and volume densities when the city population grows.
Colors denote the growth categories identified in Fig.~\ref{fig:VAP}, while the percentages in parentheses represent the proportion of cities belonging to each category.
The dots on the plane represent the surface population density, $\rho_A=P/A$, whereas the anthropomorphic icons enclosed in boxes represent the volume population density, $\rho_V=P/V$. 
Surface density increases, remains approximately constant, or decreases when $\beta_A<1$, $\beta_A\approx1$, or $\beta_A>1$, respectively. 
Likewise, volume density increases, remains approximately constant, or decreases when $\beta_V<1$, $\beta_V\approx1$, or $\beta_V>1$, respectively. 
Note that this figure is purely schematic, and the numbers of dots and people shown are illustrative only and do not quantitatively represent the magnitude of the density changes within each growth category.
}
\label{fig:squares}
\end{figure}

\subsection*{Scaling growth categories}
Next, we classify the cities in Fig.~\ref{fig:VAP}g according to whether their scaling exponents ($\beta_A$ and $\beta_V$) are sublinear ($\beta<1$), statistically indistinguishable from linear ($\beta\approx1$), or superlinear ($\beta>1$) at the 99\,\% confidence level. In principle, this classification yields nine possible combinations.
Two of the nine theoretical categories ($\beta_A<1, \beta_V>1$ and $\beta_A \approx 1,\beta_V>1$) have zero occurrences in our dataset, and nearly 90\,\% of the cities fall into only three categories: both exponents sublinear ($\beta_A,\beta_V<1$), both approximately linear ($\beta_A,\beta_V\approx1$), or both superlinear ($\beta_A,\beta_V>1$), with the sublinear category being by far the most prevalent.
Importantly, the vast majority of cities in these three dominant categories satisfy $\beta_V<\beta_A$, and therefore exhibit $\beta_h<0$, implying that the average building height decreases as population increases.
Figure~\ref{fig:VAP}h presents the geographic distribution of the different scaling categories. While most of the categories are observed worldwide, superlinear cities are particularly concentrated in India and neighboring countries. In contrast, Chinese cities, which have been experiencing rapid urbanization over recent decades, are distributed across several scaling categories.

The implications of the different growth categories remain to be understood. Figure~\ref{fig:squares} illustrates how the scaling exponents determine the evolution of the surface population density (i.e., population per unit area $\rho_A$, hereafter referred to as \emph{population density}) and the volume population density $\rho_V$ (i.e., population per unit volume).
We begin with the surface population density, $\rho_A$, represented by the dots on the plane in this figure. When $\beta_A<1$ (blue), the population grows faster than the urban area, resulting in an increasing population density. If $\beta_A\approx1$ (orange, red), population and area increase at approximately the same rate, and the population density remains nearly constant. Finally, when $\beta_A>1$ (green, violet), the urban area expands faster than the population, leading to a decrease in population density.
A similar interpretation applies to the volume population density, $\rho_V$, represented by the ``people'' enclosed within a box in the figure. When $\beta_V<1$ (blue, red), the population grows faster than the building volume, resulting in an increasing volume population density. If $\beta_V\approx1$ (orange, violet), population and volume increase at approximately the same rate, and the volume population density remains nearly constant. Finally, when $\beta_V>1$ (green), building volume expands faster than population, leading to a decrease in volume population density.

Returning to the global map in Fig.~\ref{fig:VAP}h, we find that the most common category (blue), characterized by increasing surface and volume population densities, is distributed across all continents. The second most prevalent category (orange), in which both densities remain approximately constant, is likewise widespread. In contrast, the third most common category (green), characterized by decreasing surface and volume population densities, is concentrated primarily in the Indian subcontinent.
It is worth noting, however, that Indian cities have relatively high population densities, suggesting that the observed decrease in density may reflect changes relative to already dense urban environments.
Again, in the vast majority of cases, including those with $\beta_A\approx1$ and $\beta_V\approx1$, we find $\beta_h<0$, indicating that the average building height decreases over time.

\begin{figure}[htb]
\centering
\includegraphics[width=1\linewidth]{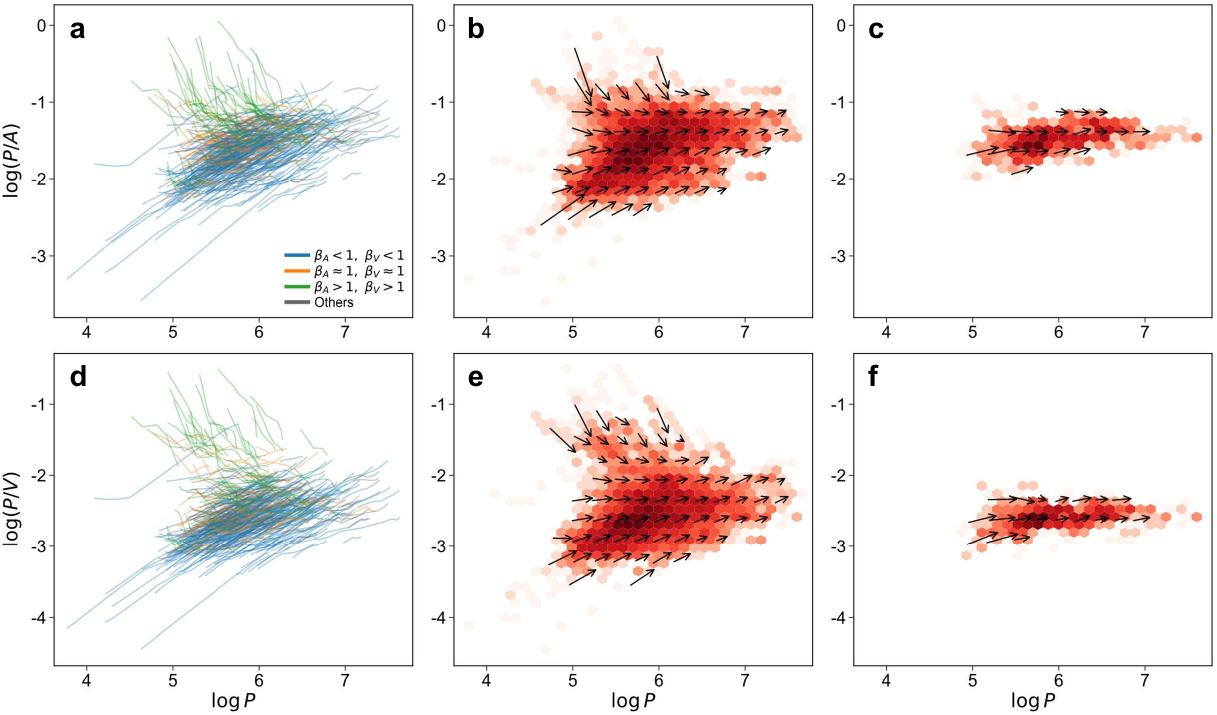}
\caption{Convergence of population densities. 
Top row panels show the surface population density as a function of population, whereas the bottom row panels show the volume population density as a function of population; both axes are logarithmic in all panels.
(a) Temporal trajectories of all cities in the $\rho_A=P/A$ vs.\ $P$ plane. The trajectory colors correspond to the growth categories defined in Figs.~\ref{fig:VAP} and~\ref{fig:squares}. 
(b) Distribution of observations in the $\log_{10}(P/A)$-$\log_{10}(P)$ plane, estimated using hexagonal binning, with darker shades indicating a higher number of observations. The arrows represent the average displacement over a five-year interval. The trajectories indicate convergence toward a characteristic surface population density in the range $0.01<P/A<0.1$ inhabitants per square meter.
(c) Same as panel~b, but restricted to cities in China.
Panels~d--f are the same as panels~a--c but for the volume population density, $\rho_V=P/V$.
}
\label{fig:vectors}
\end{figure}

\subsection*{The Role of Population Density}
To investigate the factors that determine a city's growth category, we examine the role of population density. Figure~\ref{fig:vectors} shows the evolution of surface and volume population densities as a function of population, with both axes displayed on logarithmic scales. Figures~\ref{fig:vectors}a-b, and~\ref{fig:vectors}d-e,  show the trajectories of all cities in the planes $\log(P/A)$ versus $\log(P)$ and $\log(P/V)$ versus $\log(P)$, respectively.
A clear pattern emerges. Cities with initially high population densities tend to exhibit negative slopes, indicating that their densities decrease as they grow. In contrast, cities with initially low population densities exhibit positive slopes, implying that their densities increase with population growth.
The colors are the same as in Fig.~\ref{fig:VAP}h and correspond to the growth categories illustrated in Fig.~\ref{fig:squares}. As expected, cities in the sublinear category (blue) become denser as they grow, whereas cities in the superlinear category (green) exhibit a gradual decrease in population density.

Taken together, these trajectories, indicated by arrows, suggest a convergence toward a characteristic population density in the range $0.01<P/A<0.1$ inhabitants per square meter, as shown more clearly in Fig.~\ref{fig:vectors}b and c.
Because urban area and building volume are strongly correlated (Fig.~\ref{fig:VAP}), the volume population density follows the same qualitative pattern. As cities grow, they tend to converge toward a characteristic volume population density in the range $0.001<P/V<0.01$ inhabitants per cubic meter.

In summary, cities with initially high population densities tend to reduce their densities as they grow, whereas cities with initially low population densities become denser over time. These opposing trends suggest a convergence toward a characteristic population density during urban growth.
Figures~\ref{fig:vectors}c and~f reproduce the same analysis using only cities in China. The qualitative behavior remains unchanged, although the observed density range is naturally narrower.

\subsection*{Future Scenarios}
\begin{figure}[H]
\centering
\includegraphics[width=1\linewidth]{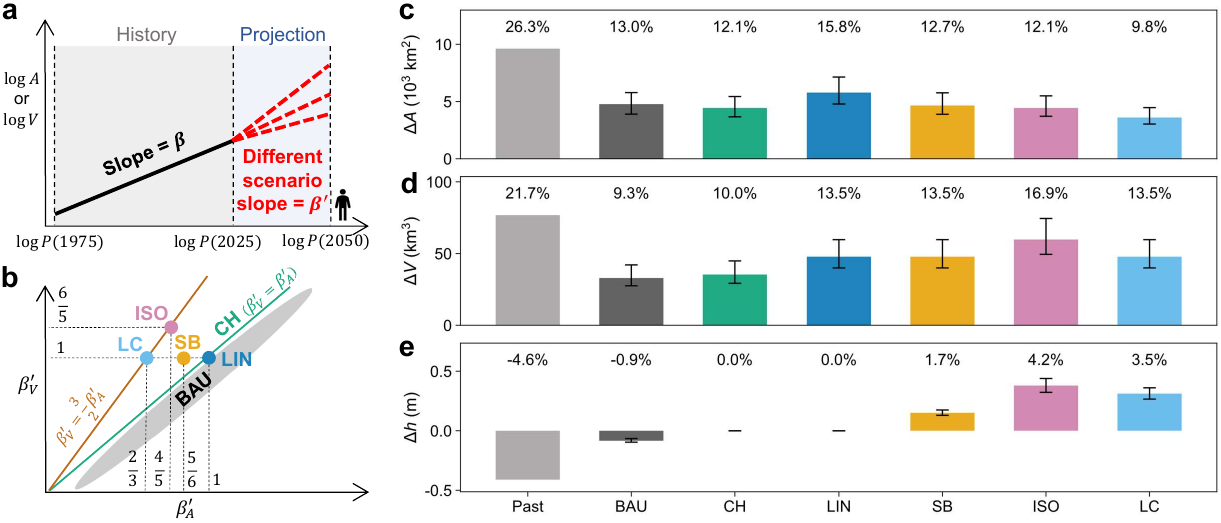}
\caption{
Projection of global increases in urban area, building volume, and average building height under different future scaling scenarios derived from historical exponents.
(a) Schematic illustration of the projection framework. 
The projected global population for 2050 \cite{wang2022projecting}, estimated from the SSP scenarios,  is used as the driver of urban growth. Combined with the scenario-specific scaling exponents $\beta_A'$ and $\beta_V'$, it allows us to estimate the future increases in urban area, building volume, and average building height.
(b) Overview and schematic representation of the future scaling scenarios in the $\beta_V'$--$\beta_A'$ plane. The scaling scenarios correspond to those listed in Table~\ref{tab:scenarios}: BAU (business as usual), CH (constant height), LIN (linear), SB (Schl\"apfer--Bettencourt), ISO (isotropic), and LC (Lemoy--Caruso). In the BAU scenario, each city retains its historical scaling exponents.
The lower growth in the BAU case compared to the past can be attributed to the smaller upcoming population increase.
In the CH scenario, each city is assigned a new pair of exponents lying on the diagonal ($\beta_V'=\beta_A'$), while preserving the mean of its historical exponents. In the remaining scenarios (LIN, SB, ISO, and LC), all cities are assigned the same pair of scaling exponents, indicated by the filled circles.
(c)--(e) Estimated growth by 2050 obtained using the projection framework described in Section~\ref{sec:projection_calculos}. The bars show the mean increase in urban area (c), building volume (d), and average building height (e) across the five SSP 
\cite[e.g.,][]{dellink2017long}. 
Error bars indicate the range between the lowest and highest projections among the five scenarios. The percentages above the bars are relative to the 2025 values, except for the light-gray bars (historical), which are expressed relative to 2000. Note that, consistent with our previous analysis, only cities exhibiting continuous population growth are included in this assessment.
}
\label{fig:predict}
\end{figure}

Finally, we investigate future scaling scenarios for the growth of urban building area, building volume, and average building height. These scenarios are described in detail in the Materials and Methods section (Sec.~\ref{sssec:scenarios}) and summarized in Table~\ref{tab:scenarios} and in Fig.~\ref{fig:predict}-a-b. Besides providing a framework for projecting future urban development, they also facilitate comparisons between our results and previous theoretical and empirical studies.

\begin{table}[htb]
\centering
\begin{tabular}{lllll}
\hline
area exponent & volume exponent & height exponent & scenario designation & abbreviation \\
\hline
$\beta_A'=\beta_A$ & $\beta_V'=\beta_V$ & $\beta_h'=\beta_V-\beta_A$ & business as usual & BAU \\
$\beta_A'=(\beta_A+\beta_V)/2$ & $\beta_V'=(\beta_A+\beta_V)/2$ & $\beta_h'=0$ & constant height & CH \\
$\beta_A'=1$ & $\beta_V'=1$ & $\beta_h'=0$ & linear (also const.\ height) & LIN \\
$\beta_A'=5/6$ & $\beta_V'=1$ & $\beta_h'=1/6$ & Schl\"apfer-Bettencourt & SB \\
$\beta_A'=4/5$ & $\beta_V'=6/5$ & $\beta_h'=2/5$ & isotropic & ISO \\
$\beta_A'=2/3$ & $\beta_V'=1$ & $\beta_h'=1/3$ & Lemoy-Caruso (also isotropic) & LC \\
\hline
\end{tabular}
\caption{Scenarios exponents of growth scaling in terms of city area, volume, and average building height. 
For a visualization see Fig.~\ref{fig:predict}b; details can be found in the material and methods section \ref{sssec:scenarios}.
Please note that for the first two, BAU and CH, each city has its own set of exponents, while for the remaining scenarios all cities get the same exponents as specified in the table.
}
\label{tab:scenarios}
\end{table}

Between 2000 and 2025, the total urban area and building volume of the cities in our sample increased by more than 20\,\%, whereas the average building height decreased by nearly 5\,\% (Fig.~\ref{fig:predict}c--e, light-gray bars). To project future changes, we use population scenarios as drivers of urban expansion (Fig.~\ref{fig:predict}a) in combination with the proposed scaling scenarios.

Before discussing the projections, it is important to note that the population of the cities in our sample increased by approximately 55\,\% between 2000 and 2025, whereas it is projected to change by only about $-14\,\%$ to $5\,\%$ over the period 2025--2050, depending on the underlying Shared Socioeconomic Pathway (SSP) \cite{wang2022projecting}. Consequently, future urban growth is expected to be substantially more modest than in the historical period. Nevertheless, our projections still indicate appreciable increases in urban area and building volume, as we assume that existing buildings are generally not decommissioned even in cities experiencing population decline.

Figure~\ref{fig:predict}c--e presents the projected global increases in urban area, building volume, and average building height under the proposed future scaling scenarios. We first discuss the projected changes in urban area (Fig.~\ref{fig:predict}c).
If each city retains its historical scaling exponents (business as usual [BAU] scenario), the projected urban land expansion between 2025 and 2050 is approximately 13\,\%. Under the constant-height (CH) scenario, the projected increase is slightly smaller (12.1\,\%), reflecting the historical tendency toward decreasing average building height. The isotropic (ISO) scenarios and the exponents obtained from Schl\"apfer and Bettencourt (SB) yield comparable estimates, with projected area increases of approximately 12--13\,\%, close to the BAU result.
The linear (LIN) scenario produces a somewhat larger increase in urban area, consistent with the fact that most cities in our historical analysis exhibit sublinear area scaling ($\beta_A<1$). In contrast, the exponents obtained from Lemoy and Caruso (LC) result in substantially lower land consumption (9.8\,\%), owing to its relatively small area scaling exponent ($\beta_A'=2/3$).

Turning to building volume (Fig.~\ref{fig:predict}d), the BAU scenario projects an increase of approximately 9\,\% by 2050. Under the CH scenario, the projected increase is slightly larger (10\,\%). The LIN scenario yields a considerably greater increase (13.5\,\%), a value that is also obtained for the SB and LC scenarios because they share the same volume scaling exponent. Among all scenarios, the ISO scenario produces the largest increase in building volume, reaching approximately 17\,\%.
Overall, the projected increases in building volume closely reflect the assumed scaling exponents; larger volume exponents lead to greater projected growth.

Figure~\ref{fig:predict}e presents the projected changes in average building height under the different future scenarios. As expected, the constant-height (CH) and linear (LIN) scenarios leave the average building height unchanged. In contrast, the business-as-usual (BAU) scenario predicts a slight decrease, consistent with the historical trend. The remaining scenarios all result in increases in average building height, with the largest occurring under the isotropic (ISO) scenario (4.2\,\%). This represents a clear contrast with the decreasing trend observed over the historical period.

Our results indicate that assuming a linear relationship between population and either urban area or building volume leads to an overestimation of future urban growth. Likewise, the isotropic scenario (ISO) predicts substantially larger increases in building volume and average building height, reinforcing our finding that cities do not evolve isotropically. 
From a sustainability perspective, continued urban land expansion is undesirable because land is a finite resource \cite{JehlingKBR2025}. Likewise, the construction of additional buildings -- particularly taller ones -- requires large amounts of cement and steel, whose production is a major source of carbon emissions \cite{huang2018carbon}. Therefore, realistic scaling relationships are essential not only for improving projections of urban form but also for assessing the environmental consequences of future urban development.

\section{Discussion}
In summary, we characterize the four dimensions of growing cities -- two horizontal dimensions describing urban area, one vertical dimension represented by average building height -- which together with area define building volume --, and the population dimension. 
Our main findings can be summarized in three points.
First, urban growth is not isotropic. As cities grow in population, they expand more rapidly in the horizontal than in the vertical direction.
Second, the average building height decreases in the vast majority of the cities analyzed. Thus, vertical development is not only slower than horizontal expansion but, on average, effectively negative.
Third, despite their diverse growth trajectories, cities tend to converge toward characteristic population densities, both with respect to urban area and building volume.

The predominance of sublinear exponents ($\beta_A<1$ and $\beta_V<1$) suggests the existence of geometric economies of scale in urban growth.
As cities become larger, they tend to accommodate additional population more efficiently, requiring proportionally smaller increases in both land and built volume.
However, substantial variation exists across cities, revealing a diversity of urban development trajectories and indicating that the balance between horizontal expansion and vertical growth is not universal.

High-rise buildings and skyscrapers are undoubtedly iconic features of modern cities \cite{batty2018inventing}. It is also plausible that both their number and their height will continue to increase in the future \cite{auerbach2020forecasting}. Our results show, however, that -- at least over the past 50 years -- high-rise buildings have become progressively less important in accommodating urban population growth.
The pronounced incompatibility with isotropic growth reflects the fundamentally different roles of the vertical and horizontal dimensions in cities. Vertical expansion depends on elevators, structural engineering, and costly construction, whereas horizontal expansion is primarily enabled by transportation infrastructure such as streets and efficient public transport. The observed decline in average building height is likely driven by the development of low-rise neighborhoods at the urban fringe, which outpaces the construction of tall buildings in city centers; see also \cite{hou2026disparities}.
Accordingly, cities are \emph{not} simply scaled-up versions of their smaller selves. Instead, they evolve by changing their physical form, expanding outward at the expense of the surrounding non-urban hinterland.

The United Nations Sustainable Development Goal (SDG) indicator~11.3.1 is defined as the ratio of the land consumption rate to the population growth rate; see \cite[e.g.]{AkurajuPHKR2020} and references therein. When both land area and population grow exponentially, the ratio of their growth rates is closely related to the exponent of the power-law relating the two quantities \cite[e.g.]{ribeiro2026inprep}. 
In this sense, our scaling exponent $\beta_A$ is similar to the SDG indicator~11.3.1, with the important distinction that $\beta_A$ is estimated from a regression over a city's long-term growth trajectory and therefore represents its characteristic scaling behavior. 
Our results show that all growing cities with $\beta_A>1$ (green, violet, and brown in Fig.~\ref{fig:VAP}) have an SDG indicator larger than one. These cities, representing approximately 23\,\% of our sample, consume land at a faster rate than their population grows. However, our analysis also highlights a complementary aspect that is not captured by SDG~11.3.1. Approximately 18\,\% of the cities exhibit $\beta_V>1$, indicating that building volume increases faster than population. Such growth is also of concern because it implies greater demand for construction materials and infrastructure, particularly cement and steel, with associated environmental impacts.

With the purpose of illustrating that the area and volume exponents really matter, we consider different scenarios.
In particular, assuming linear development leads to estimates of future development that exceed those from simply extrapolating past growth, in both area and volume.
While the consequences are dramatic, per~se these scenarios are not normative.
Only in combination with objectives defined by decision makers do they become more or less advantageous.
E.g.\ if the goal is small land-take, then the Lemoy-Caruso type growth is preferable.
Small additional volume can be achieved by continuing as in the past.
In any case, policy instruments are required to control the growth and the resulting exponents.
Ultimately, trade-offs are necessary, balancing between compact and loose urban development -- both coming with advantages and disadvantages \cite[e.g.]{ReitemeyerRSGSR2025}.
By including volume \cite[e.g.]{esch2025outward}, compactness is not limited to population density but is now extended to the \emph{volume}.
Accordingly, with the availability of building data and results like the work at hand, a new way of thinking is also needed.

Our study also has some limitations.
First, our analysis is necessarily constrained by the quality of the underlying data. We rely on the Global Human Settlement Layer (GHSL) and restrict our analysis to buildings classified as \emph{residential}. Since identifying the use of individual buildings is inherently challenging, some degree of misclassification is unavoidable. Because our analysis is conducted at the city scale and focuses on long-term statistical patterns, spatial aggregation is expected to reduce its sensitivity to isolated classification errors.  We chose to consider residential buildings because they are the primary component of the built environment that accommodates the urban population. Moreover, the temporal volume estimates combine epoch-specific built-up surface with a common 2018 height reference, so changes in the volume-to-area ratio characterize the city-scale height composition associated with outward expansion and infill development, although low-rise-to-high-rise redevelopment involving little change in building footprint is not separately distinguished. Nevertheless, this harmonized construction provides a consistent basis for comparing long-term population–area–volume trajectories across cities in this study. A second source of uncertainty concerns the definition of city boundaries, a long-standing issue in urban science \cite[e.g.,]{sotomayor2020city,montero2021delineation}. To ensure consistency between population and built-environment data, we adopted the city definitions provided by the Urban Centre Database (UCDB), which is fully compatible with the GHSL.
Finally, our analysis is intentionally restricted to large, growing cities. Specifically, we consider only cities with an urban area exceeding 100\,km$^2$ that exhibit continuous population growth throughout the study period. These selection criteria improve the robustness of the estimated scaling exponents by excluding cities whose trajectories are dominated by shrinkage or strong fluctuations. Consequently, our conclusions should be interpreted as applying primarily to large cities experiencing persistent growth.

As an outlook, future work could further explore the dynamics of cities in the population--area--volume space to better characterize the diversity of urban growth trajectories. Such analyses could include shrinking cities, re-urbanization processes, and other stages of urban development \cite{carra2019fundamental}. For example, repeating our analysis for cities undergoing population decline would provide valuable insights into the scaling properties of depopulating urban systems. Another particularly intriguing direction concerns cities with approximately stable populations that nevertheless continue to expand spatially, a phenomenon that remains poorly understood \cite{JehlingKBR2025}.
The emergence of tall buildings and skyscrapers is, of course, linked to high land prices, especially in city centers.
Increasing the number of floors while keeping the footprint area constant allows for more intensive use, for example in terms of revenue.
Accordingly, it makes sense to examine skyscrapers from an economic perspective
\cite[e.g.]{jedwab2021comparing,ahlfeldt2022economics,ahlfeldt2026skyscraper}.
Last but not least, to the best of our knowledge, no current theory can explain the empirical patterns identified here. Existing urban scaling theories \cite{ribeiro2023mathematical} were largely developed before comprehensive building-level data became available and therefore do not explicitly account for building height, building volume, or volume population density. Developing a settlement scaling theory that incorporates both horizontal expansion and vertical development is therefore an important challenge for future research.

\section{Materials and Methods}
\label{sec:mat-methods}
\subsection{Data}
To investigate the temporal co-evolution of urban population, land use, and three-dimensional urban form, we combine multiple global datasets covering the period from 1975 to 2025. Our analysis relies primarily on the \emph{Global Human Settlement Layer} (GHSL), which provides harmonized information on urban boundaries, population distribution, building footprint area, and building volume at regular time intervals. These datasets enable the consistent characterization of urban growth trajectories across hundreds of cities worldwide. Further details on these datasets and their processing are provided below.

\subsubsection{City definitions}
We use urban definitions from the \emph{Urban Centre Database} (UCDB) of the GHSL~\cite{melchiorri2024UCDB,mariRivero2026UCDB}. These definitions identify cities based on the spatial concentration of high-density populations, enabling consistent cross-country comparisons beyond administrative or statistical boundaries.

\subsubsection{Urban population}\label{sec:urban-popu}
Historical population data from 1975 to 2025 every 5 years are obtained from the GHS-POP R2023A dataset~\cite{schiavina2023ghspop}. For each city and year, the total population is calculated by aggregating the gridded population values within the corresponding UCDB urban boundaries. 

Population projections for 2050 under different \emph{Shared Socioeconomic Pathways} (SSPs) are obtained from \citeauthor{wang2022projecting}~\cite{wang2022projecting}. Their projections were produced using a Random Forest algorithm based on the \emph{WorldPop} dataset and constrained to be quantitatively consistent with national-level SSP population totals, supporting its reliability for spatially explicit population analysis. For each city and scenario, the projected population in 2050 is calculated by aggregating the gridded population values within the corresponding UCDB urban boundaries.

\subsubsection{Urban 3D form: Area and Volume}
We use temporal built-up surface and volume estimates from the GHSL for 1975--2020, together with the short-range projection for 2025, at 5-year intervals~\cite{pesaresi2023ghsbuilts,pesaresi2023ghsbuiltv}. For each city, total residential building area ($A$, in $\text{m}^2$) is computed by aggregating the residential built-up area of all grid cells within the corresponding urban boundary. Similarly, total residential building volume ($V$, in $\text{m}^3$) is obtained by summing the residential built-up volume across all grid cells within the same boundary. Our analysis focuses on residential buildings to ensure that the resulting scaling relationships primarily reflect population accommodation demand. Residential structures constitute the dominant component of the urban building stock, accounting for approximately 90\,\% of total building area and volume. Importantly, the temporal gridded data provide spatially consistent estimates of built-up surface at five-year intervals. By integrating these estimates with a common spatially explicit height reference (based on the year of 2018), the dataset captures both outward urban expansion and the accumulation of built-up surface within existing urban areas. It thereby characterizes how the spatial redistribution of development across lower- and higher-rise locations shapes the evolution of citywide building volume and average height composition.

\subsubsection{City selection}
To reduce the influence of outliers associated with very small cities, particularly in South Asia, we include only cities with an area larger than $100\ \text{km}^2$, following previous studies~\cite{ortman2020cities,burger2022global}. 
We further restrict the sample to cities with continuously increasing population over the period 1975--2025, ensuring that the time series reflects sustained urban growth.
The final sample contains 589 cities. These cities account for more than half of the total urban population at the beginning of the study period and more than 70\,\% by 2025, see also Tab.~\ref{tab:ssp_population}.

\subsection{Methods}
\subsubsection{Scaling relations and densities}
We characterize the relationship between population size and the physical extent of cities using 
scaling relations.
In particular, we consider how the  
\emph{total building footprint area}~$A$ and the \emph{total building volume}~$V$ scale with population~$P$, 
i.e.\
\begin{equation}\label{Eq_AP}
A \sim P^{\beta_A}\, ,
\end{equation}
and
\begin{equation}\label{Eq_VP}
V \sim P^{\beta_V}\, . 
\end{equation}
Here, $\beta_A$ and $\beta_V$ are the scaling exponents relating population to urban area and building volume, respectively.
The exponent $\beta_A$ quantifies how the horizontal spatial extent of cities changes with population, whereas $\beta_V$ captures how the 
building volume increases with population.
From Eqs.~(\ref{Eq_AP}) and~(\ref{Eq_VP}), it is possible to derive the scaling relationship between building volume and urban area,
$V \sim A^{\frac{\beta_V}{\beta_A}}$.
The exponent $(\beta_V/\beta_A)$ characterizes how built volume increases with urban area and therefore quantifies the relative contribution of vertical development to urban expansion.

To complement these scaling relations, we define three \emph{intensive-like} derived variables\footnote{We use the term \emph{intensive-like} to denote quantities obtained as ratios of extensive urban variables \cite{Shalizi2011}. Unlike true intensive variables in thermodynamics, these quantities may still exhibit systematic scaling with city size.}.
First, the \emph{average building height} is defined as
\begin{equation}\label{Eq_h_VA}
\bar{h} = \frac{V}{A}\, ,
\end{equation}
which represents the average verticalization of the built environment. 
Previous theoretical and empirical studies based on cross-sectional urban scaling analyses~\cite{schlapfer2015skyline, Molinero2021} have suggested that mean building height also follows a scaling relationship with population size,
\begin{equation}\label{Eq_hP}
\bar{h} \sim P^{\beta_h} \, ,
\end{equation}
where $\beta_h$ denotes the scaling exponent governing the verticalization of cities (for better readability, in the following we skip the bar and use simply $h$ for the average height).
It follows that
\begin{equation}´\label{Eq_betah}
\beta_h=\beta_V-\beta_A \, .
\end{equation}
Second, we define the \emph{surface population density},
\begin{equation}
\rho_A = \frac{P}{A}\, ,
\end{equation}
which measures the number of inhabitants per unit of urban area.
This $\rho_A$ corresponds to what is normally understood as population density in the cities context.
Third, we define the \emph{volume population density},
\begin{equation}
\rho_V = \frac{P}{V}\, ,
\end{equation}
which measures the number of inhabitants per unit of built volume.

Together, these quantities provide complementary perspectives, allowing us to characterize the horizontal expansion, vertical development, and occupation intensity of cities.

\subsubsection{Isotropic case}\label{sssec:isotropic}
It can be instructive to compare the observed city growth with a reference growth behavior.
One such reference is isotropic growth.
Isotropic means that the growth is the same in all directions.
If the city on the map grows e.g.\ 10\,\% in all directions (N-S \& E-W) -- and also the building height increases by 10\,\% (e.g.\ demolishing old ones and building taller ones), then this would be isotropic growth.

In order to obtain the respective scaling, we can imagine a soap-bubble that grows on a surface.
We idealize the bubble as half of a sphere.
The volume of our hemispherical bubble scales with the radius~$r$ following $V\sim r^3$.
Similarly, the area of its circular base scales as $A\sim r^2$.
Considering the scaling relations Eq.~(\ref{Eq_VP}) and~(\ref{Eq_AP}), we can involve the population~$P$, i.e.\ $V\sim r^3 \sim P^{\beta_V}$ and $A\sim r^2 \sim P^{\beta_A}$, respectively.
Eliminating $r$, we obtain $P^{\beta_V/3} \sim P^{\beta_A/2}$ and 
\begin{equation}\label{eq:isotropic}
\beta_V=\frac{3}{2}\beta_A \, .    
\end{equation}
This means, isotropic growth represents a special case of how volume and area grow with population.
Alternatively, one can consider the ratio $\beta_V/\beta_A$.
The case $\beta_V/\beta_A\approx 3/2$, would imply isotropic growth (if $\beta_V/\beta_A\approx 1$, then the average height remains approximately constant).
Last, we need to note that Eq.~(\ref{eq:isotropic}) disregards any potential fractal properties.

\subsubsection{Estimating scaling relations employing Principal Component Analysis (PCA)}
In this study, we analyze the scaling relationships among building volume ($V$), building area ($A$), and population ($P$) in order to estimate the scaling exponents $\beta_A$,  $\beta_V$, and $\beta_{AV}$. Traditional urban scaling analyses typically rely on Ordinary Least Squares (OLS) regression, where population is treated as the independent variable and a single urban indicator is treated as the dependent variable. While this approach might be appropriate for estimating individual pairwise relationships, it does not fully exploit the fact that $P$, $A$, and $V$ are interconnected components of the urban system.
To address this limitation, we employ Principal Component Analysis (PCA). This approach offers two main advantages. First, PCA allows the three scaling relationships to be estimated simultaneously, thereby preserving the multivariate structure of the data. In contrast, fitting each relationship separately using OLS may lead to inconsistencies among the estimated exponents because each regression is performed independently. Second, PCA is based on orthogonal deviations from the principal axis of variation and therefore treats all variables symmetrically. Unlike OLS, which requires the specification of dependent and independent variables and generally yields different estimates when the axes are exchanged, PCA does not impose a directional dependence among variables.

PCA works by identifying a new set of orthogonal axes that successively maximize the variance of the data. The first principal component (PC1) captures the largest fraction of the total variance and represents the dominant trend in the dataset. The second principal component (PC2) is orthogonal to PC1 and captures the largest remaining variation not explained by PC1. Similarly, the third principal component (PC3) is orthogonal to both PC1 and PC2 and accounts for the residual variation. Together, these three components provide a complete representation of the data in the transformed coordinate system.
In logarithmic space, the urban data points are distributed approximately along a one-dimensional manifold embedded in a three-dimensional space defined by $(\log P,\log A,\log V)$, see Fig.~\ref{fig:VAP}. The first principal component captures the dominant direction of variation in this space and can therefore be interpreted as the scaling trajectory relating population, urban area, and building volume. The conventional scaling relations, such as $A \sim P^{\beta_A}$ and $V \sim P^{\beta_V}$, correspond to projections of this three-dimensional scaling structure onto the coordinate planes. The second and third principal components quantify deviations from this dominant scaling trajectory and therefore provide information about variations in urban form that are not captured by the primary scaling relationship.
Please consult \href{https://zhangpprr.github.io/Urban-temproal-scaling-of-3D-forms/}{weblink} to inspect the interactive 3D plots.

\subsubsection{Projection scaling scenarios}\label{sssec:scenarios}
In this subsection, we describe several scenarios of future urban growth in terms of the scaling of urban area, building volume, and average building height. The aim is to examine how different relationships among the scaling exponents shape the evolution of a city's physical structure and influence its capacity to accommodate population growth. These scenarios, which will be used to investigate the future projection,  are derived either from simple theoretical assumptions or from models previously proposed in the literature.

Using the primed exponents $\beta_A'$ and $\beta_V'$ to denote future scaling exponents (Table~\ref{tab:scenarios}), while $\beta_A$ and $\beta_V$ refer to the exponents estimated from historical data, we consider the following growth scenarios:

\begin{itemize}

\item {\bf business-as-usual} (BAU): $\beta_A'=\beta_A$, \quad $\beta_V'=\beta_V$, \quad $\beta_h'=\beta_V-\beta_A$.\\
The business-as-usual scenario assumes that current growth patterns persist. Consequently, driven by population growth (Fig.~\ref{fig:predict}a), urban area and building volume continue to scale as they did in the past.

\item[]

\item {\bf constant height} (CH):
$\beta_A'= \beta_V'=(\beta_A+\beta_V)/2$, \quad $\beta_h'=0$.\\
To keep the average building height constant, the future area and volume exponents must be equal, $\beta_V'=\beta_A'$, such that the corresponding height exponent vanishes: $\beta_h'=\beta_V'-\beta_A'=0$. We assign both future exponents the mean of their historical values.

\item[]

\item {\bf linear} (LIN): $\beta_A'=1$, \quad $\beta_V'=1$, \quad $\beta_h'=0$ (constant height).\\
In the linear scenario, both scaling exponents are set to unity, i.e., $\beta_A'=\beta_V'=1$. Consequently, the corresponding height exponent is also zero, $\beta_h'=0$.

\item[]

\item {\bf Schl\"apfer-Bettencourt} (SB): $\beta_A'=5/6$, \quad $\beta_V'=1$, \quad $\beta_h'=1/6$.\\
This scenario combines the theoretical value $\beta_h\approx 1/6$ proposed by Schl\"apfer~et~al.~\cite{schlapfer2015skyline}, which means that average building height increases with population, with the theoretical value $\beta_A=5/6$ derived by Bettencourt~\cite{bettencourt2013origins} -- both based on cross-sectional analyses -- yields $\beta_V=\beta_A+\beta_h=1$. Under this scenario, urban area scales sublinearly with population, whereas building volume scales linearly, implying an increase in average building height.

\item[]

\item {\bf isotropic (ISO)}: $\beta_A'=4/5$, \quad $\beta_V'=6/5$, \quad $\beta_h'=2/5$.\\
The isotropic scenario is defined by the condition $\beta_V'/\beta_A'=3/2$ (Sec.~\ref{sssec:isotropic}). As an additional constraint, we require the average of the two exponents to be unity, i.e., $(\beta_A'+\beta_V')/2=1$. Under these conditions, urban area scales sublinearly, building volume scales superlinearly, and the average building height increases with population. A similar scenario, with $\beta_A'=2/3$, was previously discussed in \cite[e.g.,][]{ribeiro2023mathematical}; however, that study did not explicitly consider the scaling of building volume.

\item[]

\item {\bf Lemoy-Caruso}  (LC): $\beta_A'=2/3$, \quad $\beta_V'=1$, \quad $\beta_h'=1/3$  (also isotropic).\\
This scenario stems from Lemoy \& Caruso~\cite{lemoy2020evidence}, who analyze intra-urban density gradients. Their cross-sectional rescaling implies $\beta_A=2/3$ and following the logic $\beta_V=1$. 
This scenario is similar to the isotropic scenario (ISO), as both predict the exponent ratio $\beta_V/\beta_A=3/2$. From Eq.~(\ref{Eq_betah}), $\beta_h=\beta_V-\beta_A$, it follows that $\beta_h=1/3$, implying that the average building height increases with population.
Under this scenario, urban area scales sublinearly, whereas building volume scales linearly, implying an increase in average building height.

\end{itemize}

In the BAU and CH scenarios, each city has its own set of scaling exponents. In contrast, the LIN, SB, ISO, and LC scenarios assign the same exponents to all cities. Relating these scenarios to the density framework in Fig.~\ref{fig:squares}, the LIN scenario preserves both surface and volume densities (orange). The SB scenario preserves only the volume density, whereas the surface density increases (pink). Finally, the ISO scenario leads to an increase in surface density and a decrease in volume density.
It is worth noting from Fig.~\ref{fig:predict}b that all scenarios (except BAU and CH) lie on the opposite side of the diagonal from the historical exponents reported in this study. Consequently, all scenarios predict an increase in average building height, in contrast to the decreasing trend observed over the historical period.

\subsubsection{Projection under future population scenarios}\label{sec:projection_calculos}
In this section, we describe the methodology used to project urban area, building volume, and average building height from $t_0 \equiv 2025$ to $t_1 \equiv 2050$ under different SSPs (see section 
\ref{sec:urban-popu})
and scaling (see section \ref{sssec:scenarios}, Tab.~\ref{tab:scenarios} or Fig.~\ref{fig:predict}) scenarios.

For each city $i$ and SSP scenario $s$, we define the projected population growth ratio as
\begin{equation}
r_{is} = \frac{P_i(t_1|s)}{P_i(t_0)} \, ,
\end{equation}
where $P_i(t_0)$ is the population at the initial year ($t_0=2025$), and $P_i(t_1|s)$ is the projected population in the final year ($t_1=2050$) under SSP scenario $s$.
For notational simplicity, we omit the scenario index ``$s$'' in the following and write $r_i$ instead of $r_{is}$. Unless otherwise stated, all derivations apply independently to each SSP scenario.
For illustrative purposes, Tab.~\ref{tab:ssp_population} presents the total population of the cities included in this study under the different SSP scenarios.

Because the building stock typically does not shrink immediately in response to population decline, our baseline projection adopts a stock-preserving population growth factor,
\begin{equation}\label{eq:noshrink}
r_i \rightarrow \max(r_i,1)\, .
\end{equation}
Consequently, cities with projected population decline are assumed to retain their 2025 building area and volume throughout the projection period.

Given these definitions, and assuming that the future scaling relations satisfy
$A(t_1)\sim P(t_1)^{\beta_A'}$ and
$V(t_1)\sim P(t_1)^{\beta_V'}$, 
where $\beta_A'$ and $\beta_V'$
represent the projection scaling scenarios, the projected urban area is given by
\begin{equation}
A_i(t_1)=A_i(t_0) r_i^{\beta_A'}\,,
\end{equation}
while the projected building volume is
\begin{equation}
V_i(t_1)=V_i(t_0) r_i^{\beta_V'}\,.
\end{equation}
In logarithmic space, this corresponds to anchoring the projection at the observed 2025 value and extrapolating it to the SSP-specific 2050 population using the scaling exponent defined by the corresponding scenario. That is, explicitly,  

\begin{equation}
\ln \left ( \frac{A_i(t_1)}{A_i(t_0)} \right)  =  \beta_A' \ln \left ( \frac{P_i(t_1|s)}{P_i(t_0)} \right) \,,
\end{equation}
for the area,
and an analogous expression can be written for building volume.

Finally, for each SSP and each scaling scenario, we aggregate the projected changes across all cities. The total changes in urban area ($\Delta A$), building volume ($\Delta V$), and average building height ($\Delta h$) are given by

\begin{equation}
\Delta A = 
\sum_i
\left(
A_i(t_1)-A_i(t_0)
\right)\, ,
\end{equation}

\begin{equation}
\Delta V =
\sum_i
\left(
V_i(t_1)-V_i(t_0)
\right)\, ,
\end{equation}
and
\begin{equation}
\Delta h =
\sum_i
\left(
h_i(t_1)-h_i(t_0)
\right) \, ,
\end{equation}
respectively. Here, the average building height is obtained from the projected building area and volume according to $h_i(t)= V_i(t)/ A_i(t)$.

\begin{table}[ht]
\centering
\begin{tabular}{lll}
\hline
 & Total  &  \\
Time & population  & Change \\
 & (billion)  &  \\
\hline
2000 & 1.006 & -- \\
2025 & 1.565  & +55.4\,\% vs.\ 2000 \\
2050 SSP1 & 1.534  & $-2.0\,\%$ vs.\ 2025 \\
2050 SSP2 & 1.644  & $+5.1\,\%$ vs.\ 2025 \\
2050 SSP3 & 1.348  & $-13.8\,\%$ vs.\ 2025 \\
2050 SSP4 & 1.614  & $+3.2\,\%$ vs.\ 2025 \\
2050 SSP5 & 1.542 & $-1.5\,\%$ vs.\ 2025 \\
\hline
\end{tabular}
\caption{ \label{tab:ssp_population}
Total population of all cities included in the analysis for the years 2000 and 2025 (observed), and projected total population for 2050 under the different Shared Socioeconomic Pathway (SSP) scenarios.
}
\end{table}

\backmatter

\newpage

\bmhead{Acknowledgements}
This work was supported by the National Natural Science Foundation of China (72242102 and 72288101). 
P.\ Z.\ acknowledges support from the program of China Scholarship Council (No. 202307090047), and support from China Association for Science and Technology Young Scientific and Technological Talents Cultivation Project Doctoral Program.
F.\ L.\ R. thanks CNPq (grant numbers 403139/2021-0 and 424686/2021-0) and Fapemig (grant number APQ-00829-21 and APQ-06541-24) for financial support.
D.\ R.\ acknowledges financial support from German Research Foundation (DFG) for the projects UPon (\#451083179) and Gropius (\#511568027).
We thank N.\ Bongaerts for help with Fig.~\ref{fig:squares}.

\bmhead{Competing interests}
The authors declare no competing interests.

\bmhead{Author contribution}
P.\ Z., F.\ L.\ R., L.\ G., H.\ V.\ T.\ M.,  and D.\ R. contributed to study design, data collection, analysis, and preparation of the manuscript. 
All authors edited the manuscript. 
F.\ L.\ R., L.\ G., B.\ J., Z.\ G., M.\,B., and D.\ R. supervised the research. 
B.\ J. and Z.\ G. secure the funding. 

\bmhead{Availability}
Code and the extracted data table are available at: \url{https://github.com/peiranzhang1999-rgb/Code-for-urban-growth-study}


\end{document}